\documentclass[a4paper,10pt]{article}

 \usepackage[cp1251]{inputenc}

\usepackage{cite,amsmath,amsfonts,amsthm,fullpage}
\usepackage{youngtab}

\newcommand{\bpow}{\mathbf{p}}

\theoremstyle{plain}

\newtheorem{Theorem}{Theorem}
\newtheorem{Lemma}{Lemma}
\newtheorem{Proposition}{Proposition}
\newtheorem{Corollary}{Corollary}
\newtheorem{Remark}{Remark}

\theoremstyle{remark}

\def\tr{\mathrm {tr}}
\def\det{\mathrm {det}}

\def\bp{\begin{Proposition}}
\def\ep{\end{Proposition}}
\def\bc{\begin{Corollary}}
\def\ec{\end{Corollary}}
\def\bl{\begin{Lemma}}
\def\el{\end{Lemma}}
\def\be{\begin{equation}}
\def\ee{\end{equation}}
\def\br{\begin{Remark}\rm\small}
\def\er{\end{Remark}}
\def\brs{\begin{remarks}.\\ \rm\
\begin{enumerate}}
\def\ers{\end{enumerate}\end{remarks}}
\def\bea{\begin{eqnarray}}
\def\eea{\end{eqnarray}}

\def\tr{\mathrm {tr}}
\def\det{\mathrm {det}}

\def\&{&{\hskip -20pt}}

\newcount\YDcount\YDcount=0
\def\YDsize{10pt}

\def\YD#1{%
\ifnum#1=0
 \ifnum\YDcount=0 \ifx\varnothing\undefined\emptyset\else\varnothing\fi
 \else\vskip1.4pt\egroup\YDcount=0\fi
\else
 \ifnum\YDcount=0 \YDcount=1\vcenter\bgroup\vskip1pt
 \else\nointerlineskip\fi
 \vbox{\hrule\hbox{\vrule height\YDsize
 \loop\hskip\YDsize\vrule\ifnum\YDcount<#1\advance\YDcount1\repeat}\hrule
 \kern-0.4pt}\expandafter\YD
\fi}

\usepackage[usenames,dvipsnames]{color}
\usepackage{ulem}

\begin{document}

\author{ Aleksander Yu.
Orlov\thanks{Institute of Oceanology, Nahimovskii Prospekt 36,
Moscow 117997, Russia, and National Research University Higher School of Economics,
International Laboratory of Representation
Theory and Mathematical Physics,
20 Myasnitskaya Ulitsa, Moscow 101000, Russia, email: orlovs@ocean.ru
}}
\title{Links between quantum chaos and counting problems}

\maketitle

\begin{abstract}

We show that Hurwitz numbers may be generated by certain correlation functions which appear in
quantum chaos.

\end{abstract}

\bigskip



$\,$

First, in short we present two different topics: Hurwitz numbers which appear in counting of branched covers of
Riemann and Klein surfaces, and the study of spectral correlation functions of products of random matrices which
belong to independent (complex) Ginibre ensembles.

There are a lot of studies on extracting information about Hurwitz numbers, on the one hand side, from integrable systems,
as it was done in
\cite{Okounkov-2000}, \cite{Okounkov-Pand-2006}, \cite{Goulden-Jackson-2008} and further developed in \cite{MM1}, \cite{MM2}, 
\cite{AMMN-2011},\cite{AMMN-2014},\cite{Harnad-2014},\cite{NO-2014},\cite{HO-2014},\cite{Zog},\cite{Carrel},
\cite{DuninKazarian---2013},
\cite{NO-LMP} (see also reviews \cite{Harnad-overview-2015} and \cite{KazarianLando}) 
and from matrix integrals \cite{MelloKochRamgoolam}, \cite{GGPN},\cite{KZ} on the other hand.
(Actually the point that there is a special family of tau functions which were introduced in \cite{KMMM} and in \cite{OS-2000}
and studied in \cite{OS-TMP},\cite{HO-Borel},\cite{O-Acta},\cite{HO-2003},\cite{O-2004-New},\cite{OShiota-2004},
\cite{HO-2006} where links 
with matrix models were written down which describes perturbation series
in coupling constants of a number of matrix models, and these very tau functions, called hypergeometric 
ones, count also special types of Hurwitz numbers.
This article is based on \cite{NO-2014}, \cite{O-2004-New} and \cite{O-TMP-2017} and it was the content of
my talk in Bialowzie meeting ``XXXVI Workshop in Geometric Method in Physics, 3-8 June 2017''.
In the last paper we put known results in quantum chaos \cite{Ak1},\cite{Ak2},\cite{AkStrahov},
The results of our the work should be compared to ones obtained in \cite{Kazakov}, \cite{Alexandrov} and 
\cite{ChekhovAmbjorn}.

\section{Counting of branched covers}

Let us consider a connected compact surface without boundary $\Omega$ and a branched covering $f:\Sigma\rightarrow\Omega$
by a connected or non-connected surface $\Sigma$. We will consider a covering $f$ of the degree $d$. It means that the
preimage $f^{-1}(z)$ consists of $d$ points $z\in\Omega$ except some finite number of points. This points are called
\textit{critical values of $f$}.

Consider the preimage $f^{-1}(z)=\{p_1,\dots,p_{\ell}\}$ of $z\in\Omega$. Denote by $\delta_i$ the degree of $f$ at $p_i$. It
means that in the neighborhood of $p_i$ the function $f$ is homeomorphic to $x\mapsto x^{\delta_i}$. The set 
$\Delta=(\delta_1\dots,\delta_{\ell})$
is the partition of $d$, that is called \textit{topological type of $z$}.

For a partition $\Delta$ of a number $d=|\Delta|$ denote by $\ell(\Delta)$ the number of the non-vanishing parts ($|\Delta|$ and
$\ell(\Delta)$ are called the weight and the length of $\Delta$, respectively). We denote a partition and its Young diagram by 
the same letter. Denote by $(\delta_1,\dots,\delta_{\ell})$ the Young diagram with rows of length $\delta_1,\dots,\delta_{\ell}$
and corresponding partition of $d=\sum \delta_i$.

Fix now points $z_1,\dots,z_{\textsc{f}}$ and partitions $\Delta^{(1)},\dots,\Delta^{(\textsc{f})}$ of $d$. Denote by
\[\widetilde{C}_{\Omega (z_1\dots,z_{\textsc{f}})} (d;\Delta^{(1)},\dots,\Delta^{(\textsc{f})})\]
the set of all branched covering $f:\Sigma\rightarrow\Omega$ with critical points $z_1,\dots,z_{\textsc{f}}$ of topological
types  $\Delta^{(1)},\dots,\Delta^{(\textsc{f})}$.

Coverings $f_1:\Sigma_1\rightarrow\Omega$ and $f_2:\Sigma_2\rightarrow\Omega$ are called isomorphic if there exists an
homeomorphism $\varphi:\Sigma_1\rightarrow\Sigma_2$ such that $f_1=f_2\varphi$. Denote by $\texttt{Aut}(f)$  the group of
automorphisms of the covering $f$. Isomorphic coverings have isomorphic groups of automorphisms of degree $|\texttt{Aut}(f)|$.

Consider now the set $C_{\Omega (z_1\dots,z_{\textsc{f}})} (d;\Delta^{(1)},\dots,\Delta^{(\textsc{f})})$ of isomorphic classes
in $\widetilde{C}_{\Omega (z_1\dots,z_{\textsc{f}})} (d;\Delta^{(1)},\dots,\Delta^{(\textsc{f})})$. This is a finite set.
The sum
\be\label{Hurwitz-number-geom-definition}
H^{\textsc{e},\textsc{f}}(d;\Delta^{(1)},\dots,\Delta^{(\textsc{f})})=
\sum\limits_{f\in C_{\Omega (z_1\dots,z_{\textsc{f}})}(d;\Delta^{(1)},\dots,
\Delta^{(\textsc{f})})}\frac{1} {|\texttt{Aut}(f)|}\quad,
\ee
don't depend on the location of the points $z_1\dots,z_{\textsc{f}}$ and is called \textit{Hurwitz number}.
Here $\textsc{f}$ denotes the number of the branch points, and $\textsc{e}$ is the Euler characteristic of the base surface.

In case it will not produce a confusion we admit 'trivial' profiles $(1^d)$ among $\Delta^1,\dots,\Delta^{\textsc{f}}$ in
(\ref{Hurwitz-number-geom-definition})
keeping the notation $H^{\textsc{e},\textsc{f}}$ though the number of critical points now is less than $\textsc{f}$.

In case we count only connected covers $\Sigma$ we get the \textit{connected} Hurwitz numbers 
$H_{\rm con}^{\textsc{e},\textsc{f}}(d;\Delta^{(1)},\dots,\Delta^{(\textsc{f})})$.

\vspace{1ex}

The Hurwitz numbers arise in different fields of mathematics: from algebraic geometry to integrable systems. They are well
studied for orientable $\Omega$. In this case the Hurwitz number coincides with the weighted number of holomorphic branched
coverings of a Riemann surface $\Omega$ by other Riemann surfaces, having critical points $z_1,\dots,z_\textsc{f}\in\Omega$ of
the topological types $\Delta^{(1)},\dots,\Delta^{(\textsc{f})}$ respectively. The well known isomorphism between Riemann
surfaces and complex algebraic curves gives the interpretation of the Hurwitz numbers as the numbers of morphisms of
complex algebraic curves.

Similarly, the Hurwitz number for a non-orientable surface $\Omega$ coincides with the weighted number of the dianalytic
branched coverings of the Klein surface without boundary by another Klein surface and coincides with the weighted number
of morphisms of real algebraic curves without real points \cite{AG,N90,N2004}. An extension of the theory to all Klein surfaces
and all real algebraic curves leads to Hurwitz numbers for surfaces
with boundaries may be found in \cite{AN,N}.

Riemann-Hurwitz formula related the Euler characteristic of the base surface $\textsc{e}$ and the Euler characteristic of
the $d$-fold cover $\textsc{e}'$ as follows:
\be\label{RH}
 \textsc{e}'= d\textsc{e}+\sum_{i=1}^{\textsc{f}}\left(\ell(\Delta^{(i)})-d\right)=0
\ee
where the sum ranges over all branch points $z_i\,,i=1,2,\dots$ with ramification profiles given by partitions $\Delta^i\,,i=1,2,\dots$
respectively, and $\ell(\Delta^{(i)})$ denotes the length of the partition $\Delta^{(i)}$ which is equal to the number of
the preimages $f^{-1}(z_i)$ of the point $z_i$.

\vspace{1ex}
{\bf Example 1}.
Let $f:\Sigma\rightarrow\mathbb{CP}^1$ be a covering without critical points.
Then, each $d$-fold cover is a union of $d$ Riemann spheres: $\mathbb{CP}^1 \coprod \cdots \coprod \mathbb{CP}^1$, then
$\deg f =d!$ and $H^{2,0}(d)=\frac{1}{d!}$

\vspace{1ex}
{\bf Example 2}.
Let $f:\Sigma\rightarrow\mathbb{CP}^1$ be a $d$-fold covering with two critical points with the profiles 
$\Delta^{(1)}=\Delta^{(2)}=(d)$.
(One may think of $f=x^d$). Then $H^{2,2}(d;(d),(d))=\frac 1d$. Let us note that $\Sigma$ is connected in this case 
(therefore $H^{2,2}(d;(d),(d))=H_{\rm con}^{2,2}(d;(d),(d)) $)
and its Euler
characteristic $\textsc{e}'=2$.

\vspace{1ex}
{\bf Example 3}.
 The generating
function for the Hurwitz numbers $H^{2,2}(d;(d),(d))$ from the previous Example may be writen as
$$
F(h^{-1}\bpow^{(1)},h^{-1}\bpow^{(2)}):=\, h^{-2}\sum_{d>0}\, H_{\rm con}^{2,2}(d;(d),(d)) p_d^{(1)}p_d^{(2)}=
h^{-2}\sum_{d>0} \frac 1d p_d^{(1)}p_d^{(2)}
$$ 
Here $\bpow^{(i)}=(p_1^{(i)},p_2^{(i)},\dots),\,i=1,2$ are two sets of formal parameters. The powers of the auxilary
parameter $\frac 1h$ count the Euler characteristic of the cover $\textsc{e}'$ which is 2 in our example.
Then thanks to the known general statement about the link between generating functions of "connected" and "disconnected"
Hurwitz numbers (see for instance \cite{ZL}) one can write down the generating function for the Hurwitz numbers for covers with
two critical points,
$H^{2,2}(d;\Delta^{(1)},\Delta^{(2)})$, as follows:
\[
\tau(h^{-1}\bpow^{(1)},h^{-1}\bpow^{(2)})=e^{F(h^{-1}\bpow^{(1)},h^{-1}\bpow^{(2)}) } =
\]
\be\label{E=2,F=2Hurwitz}
e^{h^{-2}\sum_{d>0} \frac 1d p_d^{(1)}p_d^{(2)}}\,
= \,
\sum_{d\ge 0} \sum_{\Delta^{(1)},\Delta^{(2)}}
 H^{2,2}(d;\Delta^{(1)},\Delta^{(2)}) \,h^{-\textsc{e}'} \bpow^{(1)}_{\Delta^{(1)}}\bpow^{(2)}_{\Delta^{(2)}}
\ee
where $\bpow^{(i)}_{\Delta^{(i)}}:=p^{(i)}_{\delta^{(i)}_1}p^{(i)}_{\delta^{(i)}_2}p^{(i)}_{\delta^{(i)}_3}\cdots$, $i=1,2$
and where $\textsc{e}'=\ell(\Delta^{(1)}) + \ell(\Delta^{(2)})$ in agreement with (\ref{RH}) where we put $\textsc{f}=2$.
From (\ref{E=2,F=2Hurwitz}) it follows that the profiles of both critical points
coincide, otherwise the Hurwitz number vanishes. Let us denote this profile $\Delta$, 
and $|\Delta|=d$ and from the last equality we get
$$
H^{2,2}(d;\Delta,\Delta) = \frac {1}{z_{\Delta}}
$$
Here
\be
z_\Delta\,=\,\prod_{i=1}^\infty \,i^{m_i}\,m_i!
\ee
where $m_i$ denotes the number of parts equal to $i$ of the partition $\Delta$ (then the partition $\Delta$ is often
denoted by $(1^{m_1}2^{m_2}\cdots)$).

\vspace{1ex}
{\bf Example 4}.
Let $f:\Sigma\rightarrow\mathbb{RP}^2$ be a covering without critical points.
Then, if $\Sigma$ is connected, then $\Sigma=\mathbb{RP}^2$,
$\deg f=1$\quad or $\Sigma=S^2$, $\deg f=2$. Next, if $d=3$, then
$\Sigma=\mathbb{RP}^2\coprod\mathbb{RP}^2\coprod\mathbb{RP}^2$ or $\Sigma=\mathbb{RP}^2\coprod S^2$.
Thus $H^{1,0}(3)=\frac{1}{3!}+\frac{1}{2!}=\frac{2}{3}$.

\vspace{1ex}
{\bf Example 5}.
Let $f:\Sigma\rightarrow\mathbb{RP}^2$ be a covering with a single critical point with profile $\Delta$, and $\Sigma$ 
is connected.
 Note that due to (\ref{RH}) the Euler
characteristic of $\Sigma$ is $\textsc{e}'=\ell(\Delta)$. 
(One may think of $f=z^d$ defined in the unit disc where we identify $z$ and $-z$ if $|z|=1$).
In case we cover the Riemann sphere by the Riemann sphere $z\to z^m$ we get
two critical points with the same profiles. However we cover $\mathbb{RP}^2$ by the Riemann sphere, then we have the 
composition of the
mapping $z\to z^{m}$ on the
Riemann sphere and the factorization by antipodal involution $z\to - \frac{1}{\bar z}$. Thus we have the ramification 
profile $(m,m)$
at the single critical point $0$ of $\mathbb{RP}^2$.
The automorphism group is the dihedral group of the order $2m$ which consists of rotations on $\frac{2\pi }{m}$ and 
antipodal involution
$z\to -\frac{1}{\bar z}$.
Thus we get that 
$$
H_{\rm con}^{1,1}\left(2m;(m,m)\right)=\frac{1}{2m}
$$
From (\ref{RH}) we see that $1=\ell(\Delta)$ in this case.
Now let us cover $\mathbb{RP}^2$ by $\mathbb{RP}^2$ via $z\to z^d$. From (\ref{RH}) we see that $\ell(\Delta)=1$.
For even $d$ we have the critical point
$0$, in addition each point of the unit
circle $|z|=1$ is critical (a folding), while from the beginning we restrict our consideration only on isolated critical points.
For odd $d=2m-1$ there is
the single critical point $0$, the automorphism group consists of rotations on the angle $\frac{2\pi}{2m-1}$. Thus in this case
$$
H_{\rm con}^{1,1}\left(2m-1;(2m-1)\right)=\frac{1}{2m-1}
$$

\vspace{1ex}
{\bf Example 6}.
The generating series of the connected Hurwitz numbers with a single critical point from the previous Example  is
\[
F(h^{-1}\bpow)=
 \frac{1}{h^2}\sum_{m>0} p_m^2 H_{\rm con}^{1,1}\left(2m;(m,m)\right) +
 \frac{1}{h} \sum_{m>0} p_{2m-1} H_{\rm con}^{1,1}\left(2m-1;(2m-1)\right)
\]
where $H_{{\rm con}}^{1,1}$ describes $d$-fold covering either by the Riemann
sphere ($d=2m$) or by the projective plane ($d=2m-1$). 
We get the generating function for  Hurwitz numbers with a single critical point
\[
\tau(h^{-1}\bpow)=e^{F(h^{-1}\bpow ) } =
\]
\be\label{single-branch-point'}
e^{\frac {1}{h^2}\sum_{m>0} \frac {1}{2m}p_m^2  +\frac 1h\sum_{m {\rm odd}} \frac 1m p_m }=
\sum_{d>0} 
\sum_{\Delta\atop |\Delta|=d} h^{-\ell(\Delta)} \bpow_\Delta
H^{1,a}(d;\Delta)
\ee
where $a=0$ and if $\Delta=(1^d)$, and where $a=1$  and
 otherwise. Then $H^{1,1}(d;\Delta)$ is the Hurwitz number 
describing $d$-fold covering of $\mathbb{RP}^2$ with a single
branch point of type $\Delta=(d_1,\dots,d_l),\,|\Delta|=d$ by a (not necessarily connected) Klein surface of
Euler characteristic $\textsc{e}'=\ell(\Delta)$. For instance, for $d=3$, $\textsc{e}'=1$ we get
$H^{1,1}(3;\Delta)=\frac 13\delta_{\Delta,(3)}$.
For unbranched coverings (that is for $a=0$, $\textsc{e}'=d$) we get formula (\ref{unbranched}).

\paragraph{Tau functions.}

Let us note that the expression presented in (\ref{E=2,F=2Hurwitz}), namely,
\be\label{2KP-tau-vac-Schur}
 \tau^{\rm 2KP}_1(h^{-1}\bpow^{(1)},h^{-1}\bpow^{(2)}) = 
e^{h^{-2}\sum_{d>0} \frac 1d p_d^{(1)}p_d^{(2)}}
\ee
coincides with the simplest two-component KP tau function
with two sets of higher times $h^{-1}\bpow^{(i)},\, i=1,2$, while (\ref{single-branch-point'}) may be recognized
as the simplest non-trivial tau function of the BKP hierarchy of Kac and van de Leur \cite{KvdLbispec} 
\be
 \tau^{\rm BKP}_1(h^{-1}\bpow) =
e^{\frac {1}{h^2}\sum_{m>0} \frac {1}{2m}p_m^2  +\frac 1h\sum_{m {\rm odd}} \frac 1m p_m }
\ee
written down in
\cite{OST-I}. In (\ref{E=2,F=2Hurwitz}) and in (\ref{single-branch-point'}) the higher times are rescaled as
$p_m\to h^{-1}p_m,\,m>0$ as it is common in the study of the integrable dispersionless equations where only
the top power of the 'Plank constant' $h$ is taken into account. For instance, see \cite{NZ} where the counting
of coverings of the Riemann sphere by Riemann spheres was related to the so-called Laplacian growth problem \cite{MWZ},
\cite{Z}.
About the quasiclassical limit of the DKP hierarchy see \cite{ATZ}.
The rescaling is also common for tau functions used in two-dimensional gravity where the powers of $h^{-\textsc{e}}$ 
group contributuions of surfaces of Euler characteristic $\textsc{e}$ to the 2D gravity partition function \cite{BrezinKazakov}.
In the context of the links between Hurwitz numbers and integrable hierarchies the rescaling $\bpow \to h^{-1}\bpow$ was
considered in \cite{Harnad-overview-2015} and in \cite{NO-LMP}. In our case the role similar to $h$ plays $N^{-1}$, 
where $N$ is the size of matrices in matrix integrals.

With the help of these tau functions we shall construct integral over matrices. To do this we present the variables
$\bpow^{(i)},\, i=1,2$ and $\bpow$ as traces of a matrix we are interested in. We write $\bpow(X)=\left(p_1(X),p_2(X),\dots  \right)$,
where
\be\label{p_m(X)}
p_m(X) = \tr X^m = \sum_{i=1}^N x_i^m
\ee
and where $x_1,\dots,x_N$ are eigenvalues of $X$.

In this case we use non-bold capital letters for the matrix argument and our tau functions
are tau functions of the matrix argument:
\be\label{tau_1}
\tau_1^{\rm 2KP}(X,\bpow):=\tau_1^{\rm 2KP}(\bpow(X),\bpow)=
\sum_\lambda s_\lambda(X)s_\lambda(\bpow) =e^{\tr V(X,\bpow)}=
\prod_{i=1}^N e^{\sum_{m=1}^\infty \frac{1}{m}x_i^mp_m}
\ee
where $x_i$ are eigenvalues of $X$,
where $\bpow=(p_1,p_2,\dots)$ is a semi-infinite set of parameters,
and
\be\label{tau_1^B}
 \tau_1^{\rm BKP}(X): =\tau_1^{\rm BKP}(\bpow(X)) =
 \sum_\lambda s_\lambda(X) =\prod_{i=1}^N (1-x_i)^{-1}\prod_{i<j} (1-x_ix_j)^{-1}
\ee
Here $s_\lambda$ denotes the Schur function, see Section \ref{Partitions-and-Schur-functions} in Appendix. We recall the fact
\cite{Mac} we shall need: if $X$ is $N\times N$ matrix, then
\be\label{ell<N}
s_\lambda(X)=0,\qquad {\rm if}\quad \ell(\lambda)>N
\ee
where $\ell(\lambda)$ is the length of a partition $\lambda=(\lambda_1,\dots,\lambda_\ell),\,\lambda_\ell >0$.

For further purposes we need the following spectral invariants
of a matrix $X$:
\be\label{spectral-invariant}
{\bf P}_\Delta(X):=\prod_{i=1}^\ell p_{\delta_i}(X)
\ee
where $\Delta=(\delta_1,\dots, \delta_\ell)$ is a partition and each $p_{\delta_i}$ is defined by (\ref{p_m(X)})

In our notation one can write
\be\label{tau_1-XY}
\tau_1^{\rm 2KP}(X,Y)=\tau_1^{\rm 2KP}(\bpow(X),\bpow(Y))=
\sum_{\Delta}\frac{1}{z_\Delta} {\bf P}_\Delta(X){\bf P}_\Delta(Y)
\ee

\paragraph{Combinatorial approach.} The study of the homomorphisms between the fundemental group of the base Riemann sufrace 
of genus $g$ (the Euler characterisic is resectively $\textsc{e}=2-2g$)
with 
$\textsc{f}$ marked points and the symmetric group in the context of the counting of the non-equivalent $d$-fold covering with 
given profiles 
$\Delta^{i},\,i=1,\dots,\textsc{f}$ results to the following equation (for instance, for the details, see Appendix A written by Zagier for the 
Russian edition of \cite{ZL} or works \cite{M1}, \cite{GARETH.A.JONES})
\be\label{Hom-pi-S_d-Riemann}
\prod_{j=1}^g a_jb_ja_j^{-1}b_j^{-1}X_1\cdots X_{\textsc{f}} =1
\ee
where $a_j,b_j,X_i\in S_d$ and where each $X_i$ belongs to the cycle class $C_{\Delta^i}$. Then the Hurwitz number 
$H^{2-2g,\textsc{f}}(d;\Delta^1,\dots,\Delta^\textsc{f})$ is equal to the number of solutions of equation (\ref{Hom-pi-S_d-Riemann})
divided by the order of symmetric group $S_d$ (to exclude the equivalent solutions obtained by the conjugation of all factors in
(\ref{Hom-pi-S_d-Riemann}) by elements of the group. In the geometrical approach each conjugation means the re-enumaration of $d$ sheets
of the cover).

For instance Example 3 considered above counts non-equivalent solutions of the equation $X_1X_2=1$ with given cycle classes 
$C_{\Delta^1}$ and $C_{\Delta^2}$. Solutions of this equation consist of all elements of class $C_{\Delta^1}$ and inverse elements, so 
$\Delta^2=\Delta^1=:\Delta$. The number of elements of any class $C_\Delta$ (the cardinality of $|C_\Delta|$) divided by $|\Delta|!$ 
is $1 \over z_\Delta$ as we got in the Example 3.

For Klein surfaces (see \cite{M2},\cite{GARETH.A.JONES}) instead of (\ref{Hom-pi-S_d-Riemann}) we get 
\be\label{Hom-pi-S_d-Klein}
\prod_{j=1}^g R_j^2 X_1\cdots X_{\textsc{f}} =1
\ee
where $R_j,X_i\in S_d$ and where each $X_i$ belongs to the cycle class $C_{\Delta^i}$. In (\ref{Hom-pi-S_d-Klein}),
$g$ is the so-called genus of non-orientable surface which is related to its Eular chatacteristic $\textsc{e}$ as 
$\textsc{e}=1-g$. For the projective plane ($\textsc{e}=1$) we have $g=0$, for the Klein bottle ($\textsc{e}=1$) $g=1$.

Consider unbranched covers of the torus (equation (\ref{Hom-pi-S_d-Riemann})), projective plane and Klein bottle
(\ref{Hom-pi-S_d-Klein})). In this we put each $X_i=1$ in (\ref{Hom-pi-S_d-Riemann})) and (\ref{Hom-pi-S_d-Klein})).
Here we present three pictures, for the torus ($\textsc{e}=0$), the real projective plane ($\textsc{e}=1$) and Klein bottle 
($\textsc{e}=0$) which may be obtained by the identification of square's edges. We get $aba^{-1}b^{-1}=1$ for torus,
$abab=1$ for the projective plane and $abab^{-1}=1$ for the Klein bottle.

3 pictures.

Consider unbranched coverings ($\textsc{f}=0$).
For the real projective plane we have $g=1$ in (\ref{Hom-pi-S_d-Klein}) only one $R_1=ab$. If we treat the projective plane as the unit disk
with identfied opposit points of the boarder $|z|=1$, then $R$ is related to the path from $z$ to $-z$.
For the Klein bottle ($g=2$ in (\ref{Hom-pi-S_d-Klein})) there are $R_1=ab$ and $R_2=b^{-1}$.

To avoid confisions in what follows  we will use the notion of genus and the notations $g$ only for Rieamnn surfaces, while
the notion of the Euler characteristic $\textsc{e}$ we shall use both for orientable and non-orientable surfaces.

\section{Random matrices. Complex Ginibre ensemble.}
On this subject there is an extensive literature, for instance see \cite{Ak1}, \cite{Ak2}, \cite{AkStrahov}, 
\cite{S1}, \cite{S2}.

We will consider integrals over complex matrices $Z_1,\dots,Z_n$ where the measure is defined as
\be
d\Omega(Z_1,\dots,Z_n)= \prod_{\alpha=1}^n d\mu(Z_\alpha)=c 
\prod_{\alpha=1}^n\prod_{i,j=1}^N d\Re (Z_\alpha)_{ij}d\Im (Z_\alpha)_{ij}e^{-|(Z_\alpha)_{ij}|^2}
\ee
where the integration range is $\mathbb{C}^{N^2}\times \cdots \times\mathbb{C}^{N^2}$ and where $c$ is the normalization 
constant defined via $\int d \Omega(Z_1,\dots,Z_n)=1$.

We treat this measure as the probability measure. The related ensemble is called the ensemble of $n$ independent 
complex Ginibre enesembles. 
The expectation of a quantity
$f$ which depends on entries of the matrices $Z_1,\dots,Z_n$ is defined by
$$
\mathbb{E}(f)=\int f(Z_1,\dots,Z_n) d\Omega(Z_1,\dots,Z_n).
$$

Let us introduce
the following products
\bea\label{Z}
X&:=&(Z_1 C_1) \cdots (Z_n C_n)\\ 
\label{tildeZ^*}
Y_{t}&:=& Z^\dag_n Z^\dag_{n-1} \cdots Z^\dag_{t+1} Z^\dag_1Z^\dag_2  \cdots Z^\dag_{t} ,\qquad 0<  t < n
\eea
where $Z_\alpha^\dag$ is the Hermitian conjugate of $Z_\alpha$. 
We are interested in correlation functions of spectral invariants of matrices $X$ and $Y_t$.

We denote by $x_1,\dots,x_N$ and by $y_1,\dots,y_N$  the eigenvalues of the matrices $X$ and $Y_t$, respectively.
Given partitions $\lambda=(\lambda_1,\dots,\lambda_l)$, $\mu=(\mu_1,\dots,\mu_k)$, $l,k\le N$. Let us introduce the following 
spectral invariants
\be
{\bf P}_\lambda (X)=p_{\lambda_1}(X)\cdots p_{\lambda_l} (X),\qquad
{\bf P}_\mu(Y_t)=p_{\mu_1}(Y_t)\cdots p_{\mu_k}(Y_t)
\ee
where each $p_m(X)$ is defined via (\ref{p_m(X)}).

For a given partition $\lambda$, such that $d:=|\lambda|\le N$, let us consider the spectral invariant 
${\bf P}_\lambda$ of the matrix $XY_{t}$ 
(see (\ref{spectral-invariant})). We have

\begin{Theorem}\label{Theorem1}
 $X$ and $Y_t$ are defined by (\ref{Z})-(\ref{tildeZ^*}).
 Denote $\textsc{e}=2-2g$. 
 
(A)   Let $n > t=2g \ge 0$. 
Then
 \[
 \mathbb{E}\left({\bf P}_\lambda (X Y_{2g})\right)=
 \]
 \be
 z_\lambda \sum_{\Delta^1,\dots,\Delta^{n-2g+1}\atop |\lambda|=|\Delta^{j}| =d,\, j\le n-2g+1}
H^{2-2g,n+2-2g}(d;\lambda,\Delta^{1},\dots,\Delta^{n-2g+1})P_{\Delta^{n-2g+1}}(C' C'')
\prod_{i=1}^{n-2g}  P_{\Delta^i}(C_{2g+i})
 \ee
 where
 \be\label{C'C''2g}
 C' = C_1 \cdots C_{2g-1} ,
\qquad
C''=C_2C_4\cdots C_{2g}
\ee

(B)   Let $n > t=2g+1 \ge 1$. 
Then
 \[
 \mathbb{E}\left({\bf P}_\lambda (X Y_{2g+1})\right)=
 \] 
 \be
 z_\lambda \sum_{\Delta^1,\dots,\Delta^{n-2g+1}\atop |\lambda|= |\Delta^{j}|=d,\,j\le n-2g+1}
H^{2-2g,n+2-2g}(d;\lambda,\Delta^{1},\dots,\Delta^{n-2g+1})P_{\Delta^{n-2g}}(C')P_{\Delta^{n-2g+1}}(C'')
\prod_{i=1}^{n-2g-1}  P_{\Delta^i}(C_{2g+1+i})
 \ee
 where
 \be\label{C'C''2g+1}
C'= C_1C_3 \cdots C_{2g+1} ,
\qquad
C''=C_2C_4\cdots C_{2g}
\ee 
 
\end{Theorem}

\begin{Corollary}
Let  $|\lambda|=d\le N$ as before, and let each $C_i=I_N$ ($N\times N$ unity matrix).
Then
 \be
 \frac{1}{z_\lambda}\mathbb{E}\left({\bf P}_\lambda (X Y_{2g})\right)=
 \frac{1}{z_\lambda}\mathbb{E}\left({\bf P}_\lambda (X Y_{2g+1})\right)=
  N^{nd-\ell(\lambda)} \sum_{\textsc{e}'} N^{\textsc{e}'}
 S^{\textsc{e}'}_{\textsc{e}}(\lambda)
 \ee
 where $\textsc{e}=2-2g$ and where
 \be
 S^{\textsc{e}'}_{\textsc{e}}(\lambda) :=
\sum_{\Delta^1,\dots,\Delta^{n+\textsc{e}-1}\atop 
\sum_{i=1}^{n+\textsc{e}-1}\ell(\Delta^{i}) =L}
H^{\textsc{e},n+\textsc{e}}(d;\lambda,\Delta^{1},\dots,\Delta^{n+\textsc{e}-1}),\quad L=-\ell(\lambda)+ nd +\textsc{e}' 
 \ee
is the sum of Hurwitz numbers counting all $d$-fold coverings with the following properties:  

(i) the Euler characteristic of the base surface is $\textsc{e}$ 

(ii) the Euler characteristic of the cover is $\textsc{e}'$ 

(iii) there are at most $\textsc{f}=n+\textsc{e}$ critical points

\end{Corollary}

The item (ii) in the Corollary follows from the equality ${\bf P}_\Delta(I_N)=N^{\ell(\Delta)}$ 
(see (\ref{spectral-invariant}) and (\ref{p_m(X)})) and
from the Riemann-Hurwitz relation which relates Euler characteristics of a base and a cover via branch points
profile's lengths (see (\ref{RH})):
$$
\sum_{i=1}^{n+\textsc{e}-1}\ell(\Delta^{i}) =-\ell(\lambda) +(\textsc{f}- \textsc{e})d +\textsc{e}'
$$ 
In our case $\textsc{f}- \textsc{e}=n$.


\begin{Theorem} \label{Theorem2}
$X$ and $Y_t$ are defined by (\ref{Z})-(\ref{tildeZ^*}).
  
  (A)  If $|\lambda|\neq |\mu|$ then
 $\mathbb{E}\left({\bf P}_\lambda (X) {\bf P}_\mu(Y_t)\right)=0$.

 (B)    Let $|\lambda| = |\mu| =d$ and   $n-1 > t=2g+1 \ge 1$.

Then
 \[
 \mathbb{E}\left({\bf P}_\lambda (X) {\bf P}_\mu(Y_{2g+1})\right)=
 \]
 \be
 z_\lambda z_\mu 
  \sum_{\Delta^1,\dots,\Delta^{n-2g}\atop |\lambda|= |\Delta^{j}|=d,\,j\le n-2g}
H^{2-2g,n+2-2g}(d;\lambda,\mu,\Delta^{1},\dots,\Delta^{n-2g})P_{\Delta^{n-2g-1}}(C')P_{\Delta^{n-2g}}(C_{n}C'')
\prod_{i=1}^{n-2g-2}  P_{\Delta^i}(C_{2g+1+i}) 
 \ee 
 where $C'$ and $C''$ are given by (\ref{C'C''2g}).
 
 (C) 
 Let $|\lambda| = |\mu|$ $n > t=2g \ge 0$ . 

Then
 \[
 \mathbb{E}\left({\bf P}_\lambda (X) {\bf P}_\mu(Y_{2g})\right)=
 \]
 \be
 z_\lambda z_\mu 
 \sum_{\Delta^1,\dots,\Delta^{n-2g}\atop |\lambda|=|\Delta^{j}| =d,\, j\le n-2g}
H^{2-2g,n+2-2g}(d;\lambda,\mu,\Delta^{1},\dots,\Delta^{n-2g})P_{\Delta^{n-2g+1}}(C'C_n C'')
\prod_{i=1}^{n-2g}  P_{\Delta^i}(C_{2g+i})
 \ee
 where $C'$ and $C''$ are given by (\ref{C'C''2g+1}).

\end{Theorem}

\begin{Corollary}
 Let  $|\lambda|=d\le N$ as before, and let each $C_i=I_N$.
Then
 \be
 \frac{1}{z_\lambda z_\mu}\mathbb{E}\left({\bf P}_\lambda (X){\bf P}_\lambda(Y_{2g})\right)=
 \frac{1}{z_\lambda z_\mu}\mathbb{E}\left({\bf P}_\lambda (X){\bf P}_\lambda(Y_{2g+1})\right)
 =
 \frac{1}{z_\lambda}\mathbb{E}\left({\bf P}_\lambda (X Y_{2g})\right)=
 \frac{1}{z_\lambda}\mathbb{E}\left({\bf P}_\lambda (X Y_{2g+1})\right)
 \ee

\end{Corollary}

\begin{Theorem} 
\label{Theorem3}
$X$ and $Y_t$ are defined by (\ref{Z})-(\ref{tildeZ^*}).

 (A)    Let   $n-1 > t=2g+1 \ge 0$.

Then
 \[
 \mathbb{E}\left({\bf P}_\lambda (X) \tau^{\rm BKP}_1(Y_{2g+1})\right)=
 \]
 \be
 z_\lambda 
  \sum_{\Delta^1,\dots,\Delta^{n-2g}\atop |\lambda|= |\Delta^{j}|=d,\,j\le n-2g}
H^{1-2g,n+1-2g}(d;\lambda,\Delta^{1},\dots,\Delta^{n-2g})P_{\Delta^{n-2g-1}}(C')P_{\Delta^{n-2g}}(C_{n}C'')
\prod_{i=1}^{n-2g-2}  P_{\Delta^i}(C_{2g+1+i}) 
 \ee 
 where $C'$ and $C''$ are given by (\ref{C'C''2g}).
 
 (B) 
 Let $n > t=2g \ge 0$ . 

Then
 \[
 \mathbb{E}\left({\bf P}_\lambda (X) \tau^{\rm BKP}_1(Y_{2g})\right)=
 \]
 \be
 z_\lambda z_\mu 
 \sum_{\Delta^1,\dots,\Delta^{n-2g}\atop |\lambda|=|\Delta^{j}| =d,\, j\le n-2g}
H^{1-2g,n+1-2g}(d;\lambda,\Delta^{1},\dots,\Delta^{n-2g})P_{\Delta^{n+1-2g}}(C'C_n C'')
\prod_{i=1}^{n-2g}  P_{\Delta^i}(C_{2g+i})
 \ee
 where $C'$ and $C''$ are given by (\ref{C'C''2g+1}).

\end{Theorem}

The sketch of proof.
 
The character Frobenius-type formula by Mednykh-Pozdnyakova-Jones \cite{M2},\cite{GARETH.A.JONES} 
\be\label{Hurwitz-number}
H^{\textsc{e},k}(d;\Delta^1,\dots,\Delta^{k})=\sum_{\lambda\atop |\lambda|=d}
\left(\frac{{\rm dim}\lambda}{d!} \right)^{\textsc{e}}\varphi_\lambda(\Delta^1)\cdots 
\varphi_\lambda(\Delta^k)
\ee
where 
${\rm dim}\lambda$ is the dimension of the irreducible representation of $S_d$, and
\be
\label{varphi}
\varphi_\lambda(\Delta^{(i)}) := |C_{\Delta^{(i)}}|\,\,\frac{\chi_\lambda(\Delta^{(i)})}{{\rm dim}\lambda} ,
\quad {\rm dim}\lambda:=\chi_\lambda\left((1^d)\right)
\ee
$\chi_\lambda(\Delta)$ is the character of the symmetric group $S_d$ evaluated at a cycle type $\Delta$,
and $\chi_\lambda$ ranges over the irreducible complex characters of $S_d$ (they are
labeled by partitions $\lambda=(\lambda_1,\dots,\lambda_{\ell})$ of a given weight $d=|\lambda|$). It 
is supposed that $d=|\lambda|=|\Delta^{1}|=\cdots =|\Delta^{k}|$.
$|C_\Delta |$ is the cardinality of the cycle
class $C_\Delta$ in $S_d$.

Then we use the characteristic map relation
\cite{Mac}:
\be\label{Schur-char-map}
s_\lambda(\bpow)=\frac{{\rm dim}\lambda}{d!}\left(p_1^d+\sum_{\Delta\atop |\Delta|=d } \varphi_\lambda(\Delta)\bpow_{\Delta}\right)
\ee
where $\bpow_\Delta=p_{\Delta_1}\cdots p_{\Delta_{\ell}}$ and where $\Delta=(\Delta_1,\dots,\Delta_\ell)$ is a partition whose weight
coinsides with the weight of $\lambda$: $|\lambda|=|\Delta|$. Here 
\be
{\rm dim}\lambda =d!s_\lambda(\bpow_\infty),\qquad \bpow_\infty = (1,0,0,\dots)
\ee
is the dimension of the irreducable representation of the symmetric group $S_d$. We imply that 
$\varphi_\lambda(\Delta)=0$ if $|\Delta|\neq |\lambda|$.

Then we know how to evaluate the integral with the Schur function via Lemma
 used in \cite{O-2004-New} and \cite{NO-2014}, \cite{NO-LMP}
(for instance see \cite{Mac} for the derivation). 
\bl \label{useful-relations}
Let $A$ and $B$ be normal  matrices (i.e. matrices diagonalizable by unitary transformations). Then
Below $p_{\infty}=(1,0,0,\dots)$. 
\begin{equation}\label{sAZBZ^+'}
\int_{\mathbb{C}^{n^2}} s_\lambda(AZBZ^+)e^{-\textrm{Tr}
ZZ^+}\prod_{i,j=1}^n d^2Z=
\frac{s_\lambda(A)s_\lambda(B)}{s_\lambda(p_{\infty})}
\end{equation}
and
\begin{equation}\label{sAZZ^+B'}
\int_{\mathbb{C}^{n^2}} s_\mu(AZ)s_\lambda(Z^+B) e^{-\textrm{Tr}
ZZ^+}\prod_{i,j=1}^nd^2Z= \frac{s_\lambda(AB)}{s_\lambda(p_{\infty})}\delta_{\mu,\lambda}\,.
\end{equation}
\el
To prove Theorem \ref{Theorem1} we use that we can equate the integral over $E(\tau^{\rm 2KP}(XY_y))$ using this Lemma
and (\ref{2KP-tau-vac-Schur}) and then compare it to the same integral where now we use (\ref{tau_1}).
To prove Theorem \ref{Theorem2} in the similar way we equate $E(\tau^{\rm 2KP}(X)\tau^{\rm 2KP}(Y_y))$.
To prove Theorem \ref{Theorem3} we similarly $E(\tau^{\rm 2KP}(X)\tau^{\rm 2KP}(Y_y))$ in the same way taking into acount
also (\ref{tau_1^B}).

\section*{Acknowledgements}

The work has been funded by the RAS Program ``Fundemental problems of nonlinear mechanics'' and by the Russian Academic Excellence Project '5-100'.
I thank A. Odziyevich and university of Bialystok for warm hospitality which allowed
it is accurate to write down this work. I am grateful to S. Natanzon, A. Odziyevich, J. Harnad, A. Mironov (ITEP) and
to van de Ler for various remarks concerning the questions connected with this work. Special gratitude
to E. Strakhov for the fact that he drew my attention to the works on quantum chaos devoted to the products 
of random matrices and for fruitful discussions.

\appendix

\section{Partitions and Schur functions \label{Partitions-and-Schur-functions}}

Let us recall that the characters of the unitary group $\mathbb{U}(N)$ are labeled by partitions
and coincide with the so-called Schur functions \cite{Mac}. 
A partition 
$\lambda=(\lambda_1,\dots,\lambda_n)$ is a set of nonnegative integers $\lambda_i$ which are called
parts of $\lambda$ and which are ordered as $\lambda_i \ge \lambda_{i+1}$. 
The number of non-vanishing parts of $\lambda$ is called the length of the partition $\lambda$, and will be denoted by
 $\ell(\lambda)$. The number $|\lambda|=\sum_i \lambda_i$ is called the weight of $\lambda$. The set of all
 partitions will be denoted by $\mathbb{P}$.

The Schur function labelled by $\lambda$ may be defined as  the following function in variables
$x=(x_1,\dots,x_N)$ :
\be\label{Schur-x}
 s_\lambda(x)=\frac{\det \left[x_j^{\lambda_i-i+N}\right]_{i,j}}{\det \left[x_j^{-i+N}\right]_{i,j}}
 \ee
 in case $\ell(\lambda)\le N$ and vanishes otherwise. One can see that $s_\lambda(x)$ is a symmetric homogeneous 
 polynomial of degree $|\lambda|$ in the variables $x_1,\dots,x_N$, and $\deg x_i=1,\,i=1,\dots,N$.
  
 \br\label{notation} In case the set $x$ is the set of eigenvalues of a matrix $X$, we also write $s_\lambda(X)$ instead
 of $s_\lambda(x)$.
 \er

 There is a different definition of the Schur function as quasi-homogeneous non-symmetric polynomial of degree $|\lambda|$ in 
 other variables, the so-called power sums,
 $\bpow =(p_1,p_2,\dots)$, where $\deg p_m = m$.
 
For this purpose let us introduce 
$$
 s_{\{h\}}(\mathbf p)=\det[s_{(h_i+j-N)}(\mathbf p)]_{i,j},
$$
where $\{h\}$ is any set of $N$ integers, and where
the Schur functions $s_{(i)}$ are defined by $e^{\sum_{m>0}\frac 1m p_m z^m}=\sum_{m\ge 0} s_{(i)}(\bpow) z^i$.
If we put $h_i=\lambda_i-i+N$, where $N$
is not less than the length of the partition $\lambda$, then
\begin{equation}\label{Schur-t}
 s_\lambda(\mathbf p)= s_{\{h\}}(\mathbf p).
\end{equation}

 The Schur functions defined by (\ref{Schur-x}) and by (\ref{Schur-t}) are equal,  $s_\lambda(\bpow)=s_\lambda(x)$, 
 provided the variables $\bpow$ and $x$ are related by the power sums relation
  \be
\label{t_m}
  p_m=  \sum_i x_i^m
  \ee
  
  In case the argument of $s_\lambda$ is written as a non-capital fat letter  the definition (\ref{Schur-t}),
  and we imply the definition (\ref{Schur-x}) in case the argument is not fat and non-capital letter, and
  in case the argument is capital letter which denotes a matrix, then it implies the definition (\ref{Schur-x}) with $x=(x_1,\dots,x_N)$ being
  the eigenvalues.
  
  It may be easily checked that
  \be\label{p-to-p-in-Schur}
  s_\lambda(\bpow)=(-1)^{|\lambda|}s_{\lambda^{\rm tr}}(-\bpow)
  \ee
  where $\lambda^{\rm tr}$ is the partition conjugated to $\lambda$ (in \cite{Mac} it is denoted by $\lambda^*$). The Young diagram
  of the conjugated partition is obtained by the transposition of the Young diagram of $\lambda$ with respect to its main diagonal. 
  One gets $\lambda_1=\ell(\lambda^{\rm tr})$.

\section{Matrix integrals as generating functions of Hurwitz numbers from \cite{NO-2014},\cite{NO-LMP}
\label{Matrix-integrals}}

In case the base surface is $\mathbb{CP}^1$ the set of examples of matrix integrals generating Hurwitz numbers were studied in
works \cite{Chekhov-2014},\cite{MelloKochRamgoolam},\cite{AMMN-2014},\cite{ChekhovAmbjorn},\cite{KZ},\cite{ZL},\cite{Zog}.
One can show that the perturbation series in coupling constants of these integrals (Feynman graphs) may be related to TL
(KP and two-component KP) hypergeometric tau functions. 
It actually means that these series generate Hurwitz numbers with at most two arbitrary profiles
(An arbitray profile corresponds to a certain term in the perturbation series in the coupling constants which are higher times.
The TL and 2-KP hierarchies there are two independent sets of higher times
which yeilds two critical points for Hurwitz numbers).

Here, very briefly, we will write down few generating series for the $\mathbb{RP}^2$ Hurwitz numbers.
These series may be not tau functions themselves but may be presented as integrals of tau functions of matrix argument.
(The matrix argument, which we denote by a capital letter, say $X$, means that the power sum variables $\bpow$ are specified
as $p_i=\tr X^i,\,i>0$. Then instead of
$s_\lambda(\bpow)$, $\tau(\bpow)$ we write $s_\lambda(X)$ and $\tau(X)$). If a matrix integral in examples below is a BKP tau
function then it generates Hurwitz numbers with a single arbitrary profile and all other are subjects of restrictions
identical to those in $\mathbb{CP}^1$ case mentioned above.
In all examples $V(x,\bpow):=\sum_{m>0} \frac 1m x^m p_m$. We also recall the notation  $\bpow_\infty=(1,0,0,\dots)$.
We also recall that numbers
$H^{\textsc{e},\textsc{f}}(d;\dots)$ are Hurwitz numbers only in case $d\le N$, $N$ is the size of matrices.

For more details of the $\mathbb{RP}^2$ case see \cite{NO-2014}. New development in \cite{NO-2014} with respect to
the consideration in \cite{O-2004-New} is the usage of products of matrices. 
Here we shall consider a few examples. 
All examples include the simplest BKP tau function, of matrix argument $X$,
\cite{OST-I} defined by (compare to (\ref{sum-Schurs}))
\be\label{vac-tau-BKP'}
 \tau_1^{\rm B}(X)\,:=\,\sum_\lambda \,s_\lambda(X)\,=
 e^{\frac 12 \sum_{m>0} \frac 1m\left(\tr X^m\right)^2 + \sum_{m>0,{\rm odd}}\frac{1}{m}\tr X^m}
= \frac{\det^{\frac 12}\frac{1+X}{1-X} }{\det^{\frac 12}\left( \mathbb{I}_N \otimes \mathbb{I}_N - X\otimes X\right)}
\ee
as the part of the integration measure. Other integrands are the simplest KP tau functions
$\tau_1^{\rm 2KP}(X,\bpow):=e^{\tr V(X,\bpow)}$ where  the parameters
$\bpow$ may be called coupling constants. The perturbation series in coupling constants are expressed
as sums of products of the Schur functions over partitions and are similar to the series we considered in
the previous sections.

{\bf Example 1.} The projective analog of Okounkov's generating series for double Hurwitz series as a model of normal matrices.
From the equality
\[
\left({2\pi}{\zeta_1^{-1}} \right)^{\frac 12} e^{\frac{(n\zeta_0)^2}{2\zeta_1}} e^{\zeta_0 nc+ \frac12 \zeta_1 c^2}\,
=\,
 \int_{\mathbb{R}} e^{x_i n\zeta_0 +(cx_i- \frac12 x^2_i)\zeta_1} dx_i ,
\]
 in a similar way as was done in \cite{OShiota-2004} using $\varphi_\lambda(\Gamma)=\sum_{(i.j)\in\lambda}(j-i)$,
 one can derive
\[
 e^{n|\lambda|\zeta_0}e^{\zeta_1 \varphi_\lambda(\Gamma)}\delta_{\lambda,\mu}\,=\,\textsc{k} \,
 \int  s_\lambda(M) s_\mu(M^\dag) \det \left(MM^\dag\right)^{n\zeta_0}
 e^{-\frac12 \zeta_1\tr \left( \log \left( MM^\dag\right)\right)^2} dM
\]
where $\textsc{k}$ is unimportant multiplier, where $M$ is a normal matrix with eigenvalues $z_1,\dots,z_N$ and $\log |z_i|=x_i$,
and where
$dM=\,d_*U\,\prod_{i<j}|z_i-z_j|^2\prod_{i=1}^N d^2 z_i$. Then the $\mathbb{RP}^2$ analogue of Okounkov's generating series
may be presented as the following integral
(\cite{Okounkov-2000}) may be written
\be\label{Okounkov-tau-normal-matrices-BKP}
\sum_{\lambda\atop \ell(\lambda)\le N}e^{n|\lambda|\zeta_0 +
\zeta_1 \varphi_\lambda(\Gamma)}
s_\lambda(\bpow)=\textsc{k}
 \int  e^{V(M,\bpow)}
 e^{\zeta_0 n\tr \log \left(MM^\dag\right)-\frac12 \zeta_1\left( \tr\log \left( MM^\dag\right)\right)^2}
 \tau^{\rm B}_1(M^\dag) dM
\ee
Recall that in the work \cite{Okounkov-2000} there were studied Hurwitz numbers with an arbitrary number of simple branch points
and two arbitrary profiles. In our analog, describing the coverings of the projective plane, an arbitrary profile
only one, because, unlike the Toda lattice, the hierarchy of BKP has only one set of (continuous) higher times.

A similar representation of the Okounkov $\mathbb{CP}^1$   was earlier presented in
\cite{AlexandrovZabrodin-Okounkov}.

Below we use the following notations
 \begin{itemize}
  \item $  d_*U $ is the normalized Haar measure on $\mathbb{\mathbb{U}}(N)$: $\int_{\mathbb{U}(N)}d_*U =1$

  \item $Z$ is a complex matrix
    $$
d\Omega(Z,Z^\dag)  =\,\pi^{-n^2}\,e^{-\tr \left(ZZ^\dag\right)}\,
\prod_{i,j=1}^N \,d \Re Z_{ij}d \Im Z_{ij}
  $$

  \item Let $M$ be a Hermitian matrix the measure is defined
   $$
   dM= \, \prod_{i\le j}
d\Re M_{ij} \prod_{i<j} d\Im M
  $$

 \end{itemize}

It is known \cite{Mac}
\be\label{s-s-N_lambda-1}
\int s_\lambda(Z)s_\mu(Z^\dag)\,d\Omega(Z,Z^\dag) = (N)_\lambda\delta_{\lambda,\mu}
\ee
where $(N)_\lambda:=\prod_{(i.j)\in\lambda}(N+j-i)$ is the Pochhammer symbol
related to $\lambda$. A similar relation
was used in \cite{O-Acta},\cite{HO-2006},\cite{O-2004-New},\cite{AMMN-2014},\cite{OShiota-2004}, for models of Hermitian, complex
and normal matrices.

By $\mathbb{I}_N$ we shall denote the $N\times N$ unit matrix.
We  recall that
$$ s_\lambda(\mathbb{I}_N)=(N)_\lambda s_\lambda(\bpow_\infty)\,,
\qquad s_\lambda(\bpow_\infty) = \frac{{\rm dim}\lambda}{d!},\quad d=|\lambda|$$.

{\bf Example 2. Three branch points.}
The generating function for $\mathbb{RP}^2$ Hurwitz numbers with three ramification points, having three
arbitrary profiles:
\be\label{3-points-integral}
 \sum_{\lambda,\,\ell(\lambda) \le N}
 \frac{s_\lambda(\bpow^{(1)}) s_\lambda(\Lambda) s_\lambda(\bpow^{(2)})}{\left( s_\lambda(\bpow_\infty) \right)^2}
\ee
\[
 = \,\int \,\tau^{\rm B}_1\left( Z_1 \Lambda Z_2 \right)  \,\prod_{i=1,2} \,
  e^{V(\tr Z^\dag_i,\,\bpow^{(i)})}\,d\Omega(Z_i,Z^\dag_i)
\]
If $\bpow^{(2)}=\bpow(\texttt{q},\texttt{t})$ with any given parameters $\texttt{q},\texttt{t}$, and $\Lambda=\mathbb{I}_N$
then (\ref{3-points-integral}) is the hypergeometric BKP tau function. 

{\bf Example 3. 'Projective' Hermitian two-matrix model}.
The following integral
\[
\int \tau^{\rm B}_1(c M_2)  e^{\tr V(M_1,\bpow)+\tr (M_1 M_2)}dM_1dM_2 =
\sum_{\lambda}\,c^{|\lambda|} (N)_\lambda  s_\lambda(\bpow)
\]
where $M_1,M_2$ are Hermitian matrices is an example of the hypergeometric BKP tau function.

{\bf Example 4. Unitary matrices.} Generating series for projective Hurwitz numbers with arbitrary profiles
in $n$ branch points and restricted profiles in other points:
\be\label{multimatrix-unitary-RP2'}
\int e^{\tr (c U_1^\dag \dots U_{n+m}^\dag)}
\left(\prod_{i=n+1}^{n+m} \tau^{\rm B}_1(U_i) d_*U_i \right)
\left(\prod_{i=1}^{n} \tau^{\rm KP}_1(U_i,\bpow^{(i)}) d_*U_i \right)
=
\ee
\[
\sum_{d\ge 0}c^d \left( d! \right)^{1-m} \sum_{\lambda,\, |\lambda|=
d\atop \ell(\lambda)\le N}\, \left(\frac{{\rm dim}\lambda}{d!}  \right)^{2-m}
\left(\frac{s_\lambda(\mathbb{I}_N)}{{\rm dim} \lambda} \right)^{1-m-n}
\prod_{i=1}^n \frac{s_\lambda(\bpow^{(i)})}{{\rm dim} \lambda}
\]

Here $\bpow^{(i)}$ are parameters. This series generate certain linear combination of Hurwitz numbers for base surfaces
with Euler characteristic $2-m,\,m\ge 0$. In case $n=1$ this BKP tau function may be viewed as an analogue of the generating function of
the so-called non-connected Bousquet-Melou-Schaeffer numbers
(see Example 2.16 in \cite{KazarianLando}).
In case $n=m=1$ we obtain the following BKP tau function
\[
\int \tau^{\rm B}_1(U_2)  e^{\tr V(U_1,\bpow)+\tr (cU_1^\dag U_2^\dag)}d_*U_1d_*U_2 =
\sum_{\lambda\atop \ell(\lambda)\le N}\,c^{|\lambda|}\frac{s_\lambda(\bpow)}{(N)_\lambda}
\]

{\bf Example 5. Integrals over complex matrices}.

A pair of examples. 
An analogue of Belyi curves generating function \cite{Zog},\cite{Chekhov-2014} is as follows:
\be
\sum_{l=1}^N N^l\sum_{ \Delta^{(1)},\dots,\Delta^{(n+1)}\atop \ell(\Delta^{n+1})=l} c^d
H^{\textsc{e},n+1}(d;\Delta^{(1)},\dots,\Delta^{(n+1)})
\prod_{i=1}^{n} \bpow^{(i)}_{\Delta^{(i)}}
=\sum_{\lambda}c^{|\lambda|}\frac{(d!)^{m-2}(N)_\lambda}{({\rm dim}\lambda)^{m-2}}\,
\prod_{i=1}^{n}\frac{s_\lambda(\bpow^{(i)})}{s_\lambda(\bpow_\infty)}
\ee
\be
=\int e^{\tr (cZ_1^\dag \dots Z_{n+m}^\dag)}
\left(\prod_{i=n+1}^{n+m} \tau^{\rm B}_1(Z_i) d\Omega(Z_i,Z_i^\dag) \right)
\left(\prod_{i=1}^{n} \tau^{\rm KP}_1(Z_i,\bpow^{(i)}) d\Omega(Z_i,Z_i^\dag) \right)
\ee
where $\textsc{e}=2-m$ is the Euler characteristic of the base surface.

The series in the following example generates the projective Hurwitz numbers themselves where to get rid
of the factor $(N)_\lambda$ in the sum over partitions we use mixed integration over $\mathbb{U}(N)$ and over
complex matrices:
\be
\sum_{ \Delta^{(1)},\dots,\Delta^{(n)}}\, c^d\,
H^{1,n}(d;\Delta^{(1)},\dots,\Delta^{(n)})\,
\prod_{i=1}^{n} \bpow^{(i)}_{\Delta^{(i)}}\,
=\sum_{\lambda,\,\ell(\lambda)\le N}\,c^{|\lambda|} \frac{{\rm dim}\lambda}{d!}\,
\prod_{i=1}^{n}\frac{s_\lambda(\bpow^{(i)})}{s_\lambda(\bpow_\infty)}
\ee
\be
=\,\int \tau_1^{\rm KP}(c U^\dag Z_1^\dag \cdots Z_k^\dag,\bpow^{(n)})\tau_1^{\rm B}(U)d_*U \prod_{i=1}^{n-1}
\tau_1^{\rm KP}(Z_i,\bpow^{(i)}) d\Omega(Z_i,Z_i^\dag)
\ee
Here $Z,Z_i,\,i=1,\dots,n-1$ are complex $N\times N$ matrices and $U\in\mathbb{U}(N)$. As in the previous examples
one can specify all sets $\bpow^{(i)}=\bpow(\texttt{q}_i,\texttt{t}_i),\,i=1,\dots,n$ except a single one which in 
this case has the meaning of the BKP higher times.

\end{document}